\documentclass{osa-article}

\journal{oe}

\usepackage{ulem}

\begin{document}

\title{Analytical expression of aperture efficiency affected by Seidel aberrations}

\author{Hiroaki Imada,\authormark{1,3,*} Makoto Nagai\authormark{2}}

\address{\authormark{1} Universit\'{e} Paris-Saclay, CNRS/IN2P3, IJCLab, 91405 Orsay, France . \\
\authormark{2}Advanced Technology Center (ATC), National Astronomical Observatory of Japan (NAOJ), 2-21-1, Osawa, Mitaka, Tokyo, 181-8588, Japan.\\
\authormark{3}Current address: Kavli Institute for Physics and Mathematics of the Universe (WPI), The University of Tokyo Institutes for Advanced Study, The University of Tokyo, Kashiwa, Chiba 277-8583, Japan}

\email{\authormark{*}imada@lal.in2p3.fr} 



\begin{abstract}
The effect of aberrations on the aperture efficiency has not been discussed analytically, though aberrations determine the performance of a wide field-of-view system.
Expansion of a wavefront error and a feed pattern into a series of the Zernike polynomials enables us to calculate the aperture efficiency.
We explicitly show the aperture efficiency affected by the Seidel aberrations and derive the conditions for reducing the effects of the spherical aberration and coma.
In particular, the condition for coma can reduce a pointing error.
We performed Physical Optics simulations and found that, if the Strehl ratio is higher than 0.8, the derived expression provides the aperture efficiencies with a precision of $ < 2 \% $.
\end{abstract}

\section{Introduction}
Most of the existing radio telescopes were designed as a single-beam telescope.
Multi-pixel detectors have appeared and are still being developed \cite{Baselmans2012, Nitta2014, Datta2016, Suzuki2018}.
A wide field-of-view (FOV) radio telescope has also been obtained with ray-tracing simulation\cite{Fowler2007, Padin2008, Tsuzuki2015, Karkare2016, Kashima2018}.
However, it is not obvious whether ray tracing is a sufficient tool to design a wide FoV radio telescope because it can assess aberrations but cannot take diffraction into consideration.
Since a radio telescope generally transmits an electromagnetic wave with a long coherence length compared to the telescope size,
a wide FOV system should be evaluated and optimized in terms of both aberrations and diffraction.

The detectors for a radio telescope have an angular response, i.e., a feed pattern.
Since diffraction can be evaluated to some extent by using quasioptics, especially by the Gaussian beam theory, a feed pattern is usually approximated to a Gaussian function as a first step.
Some literature, e.g. \cite{Arimoto1974, Lowenthal1974}, has discussed the wavefront errors of  Gaussian beams.
The balanced aberrations for Gaussian beams have also been reported \cite{Mahajan1982, Mahajan1986}.
They are, however, inadequate to assess a wide FOV telescope because a signal from the universe enters it as a uniform plane wave.
The incident plane wave is partially coupled to the feed pattern, i.e. not all the incident energy is detected.
Therefore, the aperture efficiency, the ratio of the detected energy to the entering energy, is introduced as a figure of merit.
In terms of the coupling between two beams, there are some discussions in the points of view of the fiber-to-fiber or telescope-to-fiber coupling with an optical system \cite{Wagner1982,Ruilier2001,Ruilier1998,Thibault2004}.
These works investigated the analytical expressions of coupling efficiency in the case of no aberrations and evaluated the aberration effects on the coupling by numerical calculation.
It is difficult to apply their works to aperture efficiency because they implicitly took the beam truncation into account as coupling, though it should be evaluated regarding each beam independently.
That effect is known as spillover in the field of the antenna theory\cite{Goldsmith}.
The spillover efficiency is a significant factor for a radio detector, which is sensitive to ambient thermal radiation.
Moreover, since aberrations affect the coupling selectively, it is beneficial to distinguish the spillover from the coupling.

Olmi and Bolli \cite{Olmi2007} addressed the relation between wavefront error and aperture efficiency.
They pointed out that aperture efficiency was a function of a wavefront error and examined the relation between an apodized Strehl ratio and aperture efficiency.
They, however, assumed the direction of the peak gain, which cannot be determined in a practical designing phase due to aberrations.
Moreover, though a direction from the exit pupil toward a feed is intrinsically independent of an incident direction of a plane wave even in the paraxial limit, they fixed the relation between them, which was a special case.
They concluded their one-to-one correspondence between the aperture efficiency and the apodized Strehl ratio under the special assumption, which hinders us from deriving some useful conditions for the cancellation of aberrations.

Recently, Nagai and Imada \cite{Nagai2020} revealed that the aperture efficiency is determined by two spillover efficiencies and coupling efficiency.
The coupling efficiency is determined by two electric fields, an incident wave that holds the information on a wavefront error and an imaginary field illuminated by a feed.
In this paper, we will show that the coupling efficiency is analytically expressed as a function of aberrations and a feed pattern, which depend on the incident direction and the detector position, respectively.
In Section \ref{sec2}, the technical terms and assumptions used in this paper are described.
In Section \ref{sec3}, the aperture efficiency is analytically calculated and the conditions for reducing the effects of spherical aberration and coma are derived.
We verify the analytical expression using Physical Optics (PO) simulation in Section \ref{sec4}.
In Section \ref{sec5}, the precision and applications of the analytical expression are addressed.

\section{Definitions and assumptions} \label{sec2}
\subsection{System settings} \label{sec2.1}
We assume an axially symmetric optical system with an annular aperture.
The time dependence of an electromagnetic wave is assumed to be $ \exp( j \omega t ) $, where $ j = \sqrt{ -1 } $,
$ \omega = c k $, $ c $ is the speed of light, and $ k = 2 \pi / \lambda $ is the wave number.
The radius of curvature of the wavefront is positive when the wavefront is convex, as seen from the negative part of a coordinate.
An incident wave is linearly polarized and parallel to the detector polarization.

\subsection{Telescope aperture, pupil, and pupil plane} \label{sec2.2}
A telescope aperture is an opening of the first optical element that defines the energy going into the telescope.
It sometimes works as an aperture stop or otherwise the telescope has an aperture stop at a different position.
In the latter case, the incident energy is cut out by the aperture stop.
A pupil is a fundamental concept defined as an image of the aperture stop \cite{Born}.
The aperture stop determines the electromagnetic field that forms an image on the focal plane.
We refer to the infinite plane including a pupil as the pupil plane.

Once the entrance and exit pupils are considered, we can consider an equivalent system which has no optical elements in the object space and the image space.
Although the object, the entrance pupil, and  the exit pupil are sufficient for discussion of aberrations, we need the telescope aperture to define the incident energy.
Thus, we regard the telescope aperture, the entrance pupil, and the exit pupil as essential components in this paper.
The radii of the aperture, the entrance pupil, and the exit pupil are denoted by $ R_\mathrm{ ap } $, $ R_\mathrm{ en } $, and $ R_\mathrm{ ex } $, respectively.
We may consider holes at the center of each pupil if need be.
The radii of the holes are denoted by a dimensionless parameter, $ 0 \leq \varepsilon < 1 $, which is multiplied by each pupil radius.
It is assumed that the propagation from the sky to the entrance pupil is described with geometrical optics.
On the other hand, the feed beam from the focal plane to the exit pupil is assumed to be described with the Gaussian beam theory.

\subsection{Coordinates} \label{sec2.3}
\begin{figure}[!t]
\centering
\centering
\includegraphics[ width = 0.6 \hsize ]{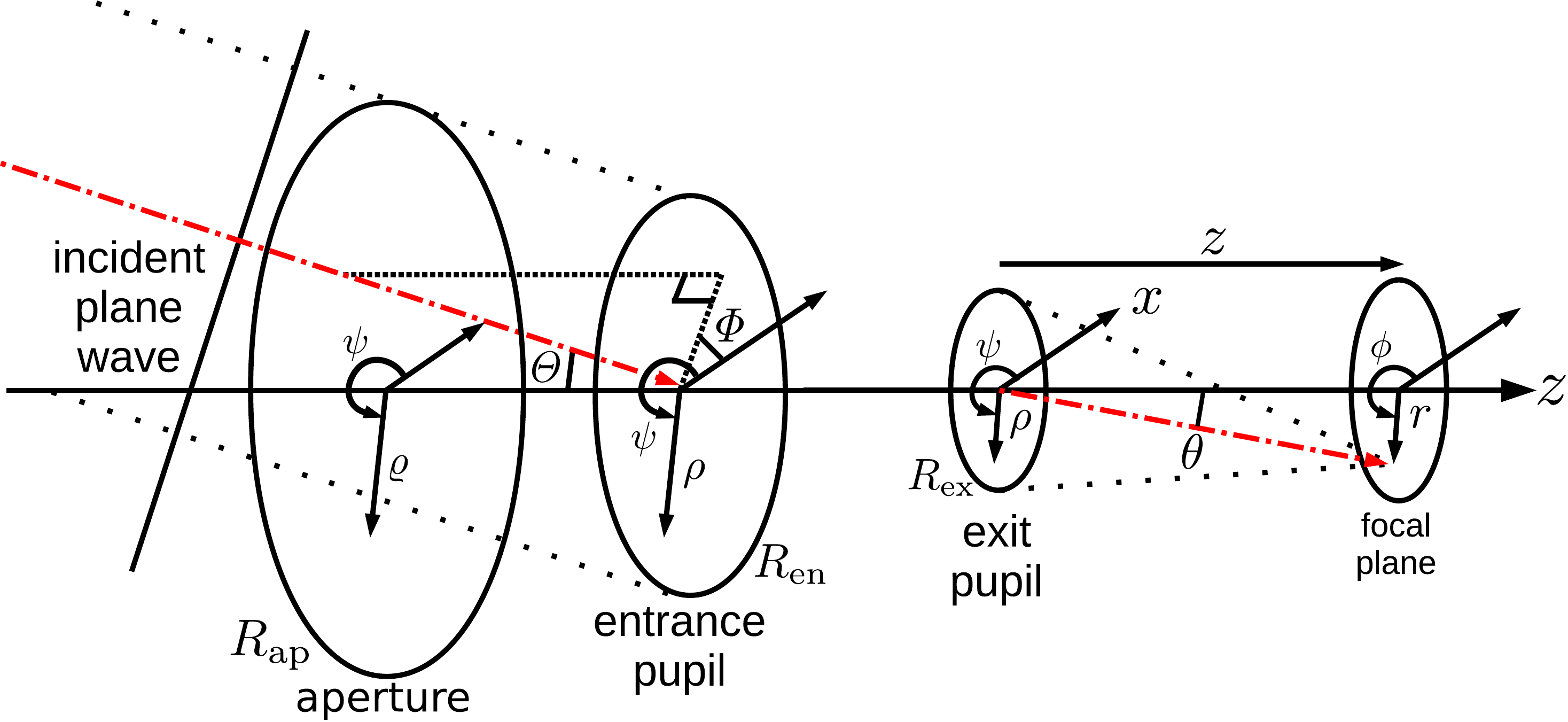}
\caption{Definition of the coordinates.}
\label{fig1}
\end{figure}

Fig.~\ref{fig1} shows the coordinates used in this paper.
An incident plane wave comes from the direction specified by an incident and azimuthal angles $ ( \varTheta, \varPhi ) $.
A cylindrical coordinates $ ( r, \phi, z ) $ are used in the image space and $ z = 0 $ is located at the exit pupil.
Another set of coordinates $ ( \rho, \psi ) $, which are common between pupils, is introduced on the pupils.
A dimensionless parameter $ \rho ~ ( \varepsilon \leq \rho \leq 1 ) $ denotes a radial distance normalized by each pupil radius and $ \psi $ denotes the azimuthal angle.
Another radial parameter on the telescope aperture, $ \varrho $, normalised by $ R_\mathrm{ ap } $, is introduced.
The focal length of the optical system is $ f $.
The incident and azimuthal angles $ ( \varTheta, \varPhi ) $ relate to a corresponding Gaussian image point\cite{Born}, $ ( r_\mathrm{ g }, \phi_\mathrm{ g }, z_\mathrm{ g } ) = ( f \tan \varTheta,  \varPhi + \pi, f R_\mathrm{ ex }/R_\mathrm{ en } ) $.
For convenience, the following vectors are introduced:
$  \boldsymbol{ p } = ( \sin \varTheta \cos \varPhi, \sin \varTheta \sin \varPhi ) $ specifying the direction of the incident wave,
$ \boldsymbol{ \varrho } = ( \varrho, \psi ) $ on the telescope aperture,
$ \boldsymbol{ \rho } = ( \rho, \psi ) $ on the pupils,
and $ \boldsymbol{ r } = ( r, \phi, z ) $ from the exit pupil center.

\subsection{Wavefront error} \label{sec2.4}
Let us focus on the wavefront of an incident wave at the exit pupil.
When no aberrations exist, it is a spherical shape whose radius equals the distance between the exit pupil center and the Gaussian image point $ \boldsymbol{ r }_\mathrm{ g } $;
however, the actual wavefront deviates from it.
Moreover, the actual wavefront is not necessarily compared to the sphere centered on $ \boldsymbol{ r }_\mathrm{ g } $.
As a consequence, we introduce a reference sphere centered on $ \boldsymbol{ r }_\mathrm{ ref } $, which is in the vicinity of $ \boldsymbol{ r }_\mathrm{ g } $, and the wavefront error $ W $ between the actual wavefront and the reference sphere along a ray.
If the actual wavefront deviates to the direction of beam propagation, $ W $ is positive.
The wavefront error $ W $ depends on the incident direction $ \boldsymbol{ p } $, position on the pupil $ \boldsymbol{ \rho } $, and reference sphere center $ \boldsymbol{ r }_\mathrm{ ref } $.
The wavefront error can be expanded into a series of the Zernike annular polynomials (ZAPs) in Eq.~(\ref{eq:zap}),
\begin{align}
W ( \boldsymbol{ p }; \boldsymbol{ \rho }; \boldsymbol{ r }_\mathrm{ ref } ) = \sum { A_n }^m ( \boldsymbol{ p }; \boldsymbol{ r }_\mathrm{ ref }; \varepsilon ) { Z_n }^m ( \boldsymbol{ \rho }; \varepsilon ), \label{eq:w}
\end{align}
where $ m $ and $ n $ are integers, and $ n - | m | \geq 0 $ is an even integer.
In this paper, we focus only on the Seidel aberrations and do not discuss random wavefront errors caused by the atmospheric fluctuation or surface roughness of the optical elements.
To consider these errors, statistical methods are needed as seen in \cite{Ruze,Fried1965, Noll1976}, and therefore, these errors are beyond the scope of this paper.
Thus, we employ ZAPs up to the third order aberrations ($ 0 \leq n + \left| m \right| \leq 4 $).

\subsection{Field at pupils} \label{sec2.5}
\subsubsection{Incident wave} \label{sec2.5.1}
An incident wave at the telescope aperture is assumed to be uniform and the electric fields at the telescope aperture and the entrance pupil can be expressed as
\begin{align}
E_\mathrm{ ap } \left( \boldsymbol{ p }; \boldsymbol{ \varrho } \right) & = \cfrac{ 1 }{ \sqrt{ \pi } R_\mathrm{ ap } }
\exp \left[ j k R_\mathrm{ ap } \varrho \sin \varTheta \cos ( \psi - \varPhi ) \right] ~ \left( 0 \leq \varrho \leq 1 \right), \label{eq:e_ap} \\
E_\mathrm{ en } ( \boldsymbol{ p }; \boldsymbol{ \rho } ) & = 
\cfrac{ 1 }{ \sqrt{ \pi } R_\mathrm{ ap } } \exp \left[ j k R_\mathrm{ en } \rho \sin \varTheta
\cos ( \psi - \varPhi ) \right] ~ ( \varepsilon \leq \rho \leq 1 ) , \label{eq:e_inc}
\end{align}
respectively. When $ \rho > 1 $ and $ \varrho > 1 $, $ E_\mathrm{ ap } = 0 $ and $ E_\mathrm{ en } = 0 $, respectively.
They are normalised by the telescope aperture area $ \pi { R_\mathrm{ ap } }^2 $.
The electric field distribution at the exit pupil is assumed to be a spherical wave with small aberrations $ W ( \boldsymbol{ p }; \boldsymbol{ \rho }; \boldsymbol{ r }_\mathrm{ ref } ) \ll \lambda $,
\begin{align}
E_\mathrm{ ex } ( \boldsymbol{ p }; \boldsymbol{ \rho } )
& = \cfrac{ R_\mathrm{ en } } { \sqrt{ \pi } R_\mathrm{ ap } R_\mathrm{ ex } } E_\mathrm{ sph } ( \boldsymbol{ \rho };
\boldsymbol{ r }_\mathrm{ ref } ) \exp \left[ j k W ( \boldsymbol{ p }; \boldsymbol{ \rho }; \boldsymbol{ r }_\mathrm{ ref } ) \right] ~ ( \varepsilon \leq \rho \leq 1 ), \label{eq:e_ex_ref}
\end{align}
where $ E_\mathrm{ sph } ( \boldsymbol{ \rho }; \boldsymbol{ r }_\mathrm{ ref } ) \coloneq
\exp \left[ - j k R_\mathrm{ ex } \rho \sin \theta_\mathrm{ ref } \cos( \psi - \phi_\mathrm{ ref } )
+ j k { R_\mathrm{ ex } }^2 \rho^2/ ( 2 z_\mathrm{ ref } ) \right] $ represents a spherical wave centered on $ \boldsymbol{ r }_\mathrm{ ref } $.
The scaling $ R_\mathrm{ en }/R_\mathrm{ ex } $ is determined according to \cite{Imada2015}.
Expanding the Taylor series of $ \exp [ j k W ( \boldsymbol{ p }; \boldsymbol{ \rho }; \boldsymbol{ r }_\mathrm{ ref } ) ] $ into ZAPs, we obtain
\begin{align}
E_\mathrm{ ex } ( \boldsymbol{ p }; \boldsymbol{ \rho } ) & = \cfrac{ R_\mathrm{ en } } { \sqrt{ \pi } R_\mathrm{ ap } R_\mathrm{ ex } }  E_\mathrm{ sph } ( \boldsymbol{ \rho }; \boldsymbol{ r }_\mathrm{ ref } ) \sum_n \frac{ \left( j k  W ( \boldsymbol{ p }; \boldsymbol{ \rho }; \boldsymbol{ r }_\mathrm{ ref } ) \right)^n }{ n ! } \nonumber \\
& = \cfrac{ R_\mathrm{ en } } { \sqrt{ \pi } R_\mathrm{ ap } R_\mathrm{ ex } } E_\mathrm{ sph } ( \boldsymbol{ \rho }; \boldsymbol{ r }_\mathrm{ ref } ) \sum_{ m, n } { B_n }^m ( \boldsymbol{ p }; \boldsymbol{ r }_\mathrm{ ref }; \varepsilon ) { Z_n }^m ( \boldsymbol{ \rho }; \varepsilon ).
\label{eq:e_ex_z}
\end{align}
When the Taylor series is taken up to the second order and the Seidel aberrations are considered, i.e. $ 0 \leq n + \left| m \right| \leq 4 $ for $ { A_n }^m $ in Eq.~(\ref{eq:w}), the coefficients $ { B_n }^m $ are explicitly given in Appendix \ref{app2}.

\subsubsection{Feed pattern} \label{sec2.5.2}
A feed pattern is often assumed to be a Gaussian function when a telescope is designed.
We here consider a feed beam whose waist size is $ w_\mathrm{ bw } $ and which is propagated from a beam waist position $ \boldsymbol{ r }_\mathrm{ bw } = \left( r_\mathrm{ bw }, \phi_\mathrm{ bw }, z_\mathrm{ bw } \right)$ toward the exit pupil center.
The angle between the optical axis and the beam propagation axis is given by $ \tan \theta_\mathrm{ bw } = r_\mathrm{ bw } / z_\mathrm{ bw } $.
We expand a feed pattern at the exit pupil plane which is given as a superposition of the Laguerre-Gaussian beam modes $ { E_{ p^\prime } }^{ q^\prime } \left( \boldsymbol{ \rho }; \boldsymbol{ r }_\mathrm{ bw }; w_\mathrm{ bw } \right) $ in Eq.~(\ref{eq:Epq_tilted}) into ZAPs,
\begin{align}
E_\mathrm{ det } ( \boldsymbol{ \rho }; \boldsymbol{ r }_\mathrm{ bw }; w_\mathrm{ bw } ) 
& = \sum_{ p^\prime, q^\prime } { D_{ p^\prime } }^{ q^\prime } { E_{ p^\prime } }^{ q^\prime }
= \frac{ E_\mathrm{ sph } 
} { R_\mathrm{ ex } } \sum_{ p, q } { C_p }^q \left( \boldsymbol{ r }_\mathrm{ bw };
\boldsymbol{ r }_\mathrm{ ref }; w_\mathrm{ bw }; \varepsilon \right) { Z_p }^q ( \boldsymbol{ \rho }; \varepsilon ), \label{eq:e_det_z}
\end{align}
where some of the arguments are omitted.
The coefficients $ { C_p }^q $ depend on $ w_\mathrm{ bw } $ and $ \boldsymbol{ r }_\mathrm{ bw } = \left( r_\mathrm{ bw }, \phi_\mathrm{ bw }, z_\mathrm{ bw } \right)$.
When $ \left| \sin \theta_\mathrm{ bw } \right| \ll 1 $ the Laguerre-Gaussian beam can be expanded into to the Taylor series up to the first order of
$ \sin \theta_\mathrm{ bw } $ and the coefficients $ { C_p }^q $ are calculated from Eq.~(\ref{eq:app_c}) in Appendix~\ref{app3},
\begin{align}
& { C_p }^q \left( \boldsymbol{ r }_\mathrm{ ref }; \boldsymbol{ r }_\mathrm{ bw }; w_\mathrm{ bw }; \varepsilon \right) = F_0
\sqrt{ 1 + p } \sum_{ p^\prime, q^\prime } { D_{ p^\prime } }^{ q^\prime } \sqrt{ \frac{ { T_\mathrm{ e } }^{ \left| q^\prime \right| } p^\prime ! }
{ \left( p^\prime + \left| q^\prime \right| \right) ! } } \nonumber \\
& \times \exp \left[ j \left( 2 p^\prime + \left| q^\prime \right| \right) \phi_0 \left( 0; z_\mathrm{ bw } \right)
- j q^\prime \phi_\mathrm{ bw } \right]
\sum_{ u = 0 }^{ p^\prime } \frac{ \left( p^\prime + \left| q^\prime \right| \right) ! \left( - T_\mathrm{ e } \right)^u }
{ \left( p^\prime - u \right) ! \left( \left| q^\prime \right| + u \right) ! u! } \nonumber \\
& \times \left\{ \delta_{ q q^\prime } \left[ \tilde{ R }_p {}^{ \left| q \right| } \left( I_{ 2 u + p + \left| q^\prime \right| } \right)
+ F_2 \tilde{ R }_p {}^{ \left| q \right| } \left( I_{ 2 u + p + \left| q^\prime \right| + 2 } \right) \right]
+ \delta_{ q q^\prime + 1 } \left[ F_1^- \tilde{ R }_p {}^{ \left| q \right| } \left( I_{ 2 u + p + \left| q^\prime \right| + 1 } \right)
\right. \right. \nonumber \\
& + \left. \left. F_3^- \tilde{ R }_p {}^{ \left| q \right| } \left( I_{ 2 u + p + \left| q^\prime \right| + 3 } \right) \right]
+ \delta_{ q q^\prime - 1 } \left[ F_1^+ \tilde{ R }_p {}^{ \left| q \right| }
\left( I_{ 2 u + p + \left| q^\prime \right| + 1 } \right) + F_3^+ \tilde{ R }_p {}^{ \left| q \right| }
\left( I_{ 2 u + p + \left| q^\prime \right| + 3 } \right) \right] \right\}, \label{eq:Cpq}
\end{align}
where
\begin{align}
\begin{split}
& F_0 = \sqrt{ \frac{ T_\mathrm{ e } } { \pi } } \frac{ \exp \left( j k z_\mathrm{ bw }
+ j \phi_0 \left( 0; z_\mathrm{ bw } \right) \right) }{  1 - \varepsilon^2 }, ~
F_2 = \frac{ j k { R_\mathrm{ ex } }^2 } { 2 } \left( \frac{ 1 } { R \left( 0; z_\mathrm{ bw } \right) } - \frac{ 1 } { z_\mathrm{ ref } } \right) \\
& F_1^\pm = \frac{ j k R_\mathrm{ ex } } { 2 } \left( \sin \theta_\mathrm{ ref } e^{ \pm j \phi_\mathrm{ ref } } - \sin \theta_\mathrm{ bw } e^{ \pm j \phi_\mathrm{ bw } } \right)
+ \frac{ T_\mathrm{ e } \sin \theta_\mathrm{ bw } e^{ \pm j \phi_\mathrm{ bw } } } { k R_\mathrm{ ex } } \left[ \frac{ \left( 2 u  + \left| q^\prime \right| + 1 \right) z_\mathrm{ bw } } { k { w_\mathrm{ bw } }^2 } \right. \\
& \left. + j \frac{ 2 p^\prime + \left| q^\prime \right| + 1 } { 2 } \right], ~
F_3^\pm = - \frac{ { T_\mathrm{ e } }^2 z_\mathrm{ bw } \sin \theta_\mathrm{ bw } e^{ \pm j \phi_\mathrm{ bw } } } { k^2 { w_\mathrm{ bw } }^2 R_\mathrm{ ex } } \left[ 1 -j \frac{ 2 z_\mathrm{ bw } } { k { w_\mathrm{ bw } }^2 } \left( 1 - \frac{ k^2 { w_\mathrm{ bw } }^2 { R_\mathrm{ ex } }^2 } { 4 T_\mathrm{ e } { z_\mathrm{ bw } }^2 } \right) \right] ,
\end{split} \label{eq:parameters}
\end{align}
\begin{align}
\begin{split}
& R ( z; z_\mathrm{ bw } ) = - ( z - z_\mathrm{ bw } ) \left( 1 + \zeta^{ -2 } \right), ~
\phi_0 ( z; z_\mathrm{ bw } ) = \tan^{ -1 } \zeta, \\
& w( z; z_\mathrm{ bw } ) = w_\mathrm{ bw } \sqrt{ 1 + \zeta^2 }, ~
\zeta \coloneq \frac{ 2 ( z - z_\mathrm{ bw } ) } { k { w_\mathrm{ bw } }^2 }, ~
T_\mathrm{ e } = \frac{ 2 { R_\mathrm{ ex } }^2 }{ w^2( 0; z_\mathrm{ bw } ) }, 
\end{split} \label{eq:Te} \\
& I_{ 2 \mu } \coloneq \int_{ \varepsilon^2 }^1 \mathrm{ d } \left( \rho^2 \right) \rho^{ 2 \mu } \exp \left( - \frac{ T_\mathrm{ e } } { 2 } \rho^2
\right)
= - \frac{ 2 \mu ! } { T_\mathrm{ e } } \sum_{ \nu = 0 }^\mu \left( \frac{ { T_\mathrm{ e } } } { 2 } \right)^{ \nu - \mu }
\frac{ \left( e^{ - T_\mathrm{ e } / 2 } - \varepsilon^{ 2 \nu } e^{ - T_\mathrm{ e } \varepsilon^2 / 2 } \right) }
{ \nu ! }. \label{eq:Ip_def}
\end{align}
The polynomial $ \tilde{ R }_p {}^{ | q | } ( I_u ) $ is defined in Appendix~\ref{app4}, e.g., Eq.~(\ref{eq:Rtilde}).
The coefficient $ F_1^\pm $ corresponds to tip-tilt and distortion of the intensity distribution caused by the calculation on a tilted plane concerning its beam axis.
The coefficient $ F_2 $ denotes defocus which can be cancelled out in principle.
The coefficient $ F_3^\pm $ is also the distortion of the intensity distribution.

\section{Aperture efficiency evaluated at pupil} \label{sec3}
Aperture efficiency $\eta_\mathrm{A}$ is factorized into entrance and exit spillover efficiencies, $ \eta_\mathrm{ sp, en } $ and $ \eta_\mathrm{ sp, ex } $, and beam coupling efficiency $\eta_\mathrm{bcp} $ \cite{Nagai2020}.
The beam coupling efficiency $ \eta_\mathrm{ bcp } $ keeps the same values among pupils.
We evaluate $ \eta_\mathrm{ bcp } $ at the entrance pupil to see the relation between the factorization and beam properties, and also at the exit pupil to relate aberrations to $ \eta_\mathrm{ A } $.
Only the fundamental-mode Gaussian beam case is considered in this section.

\subsection{Spillover efficiency}
The entrance and exit pupil spillover efficiencies are given as follows:
\begin{align}
\eta_\mathrm{ sp, en } \left( \boldsymbol{ p }; \varepsilon \right) & = \frac{ { R_\mathrm{ en } }^2 \int_\varepsilon^1 \mathrm{ d } \rho
\int_0^{ 2 \pi } \mathrm{ d } \psi \rho \left| E_\mathrm{ en } \left( \boldsymbol{ p }; \boldsymbol{ \rho } \right) \right|^2 }
{ { R_\mathrm{ ap } }^2 \int_0^1 \mathrm{ d } \varrho \int_0^{ 2 \pi } \mathrm{ d } \psi \varrho
\left|E_\mathrm{ ap } \left( \boldsymbol{ p }; \boldsymbol{ \varrho } \right) \right|^2 } \cos \varTheta = \frac{ { R_\mathrm{ en } }^2 \left( 1 - \varepsilon^2 \right) }{ { R_\mathrm{ ap } }^2 } \cos \varTheta \label{eq:sp_en}, \\
\eta_\mathrm{ sp, ex } \left( \boldsymbol{ r }_\mathrm{ bw }, w_\mathrm{ bw }; \varepsilon \right) & = \frac{ \int_\varepsilon^1 \mathrm{ d } \rho
\int_0^{ 2 \pi } \mathrm{ d } \psi \rho
\left| E_\mathrm{ det } \left( \boldsymbol{ \rho }; \boldsymbol{ r }_\mathrm{ bw }, w_\mathrm{ bw } \right) \right|^2 }
{ \int_0^\infty \mathrm{ d } \rho \int_0^{ 2 \pi } \mathrm{ d } \psi \rho
\left| E_\mathrm{ det } \left( \boldsymbol{ \rho }; \boldsymbol{ r }_\mathrm{ bw }, w_\mathrm{ bw } \right) \right|^2 }
= \exp \left( - T_\mathrm{ e } \varepsilon^2 \right) - \exp \left( - T_\mathrm{ e } \right). \label{eq:sp_ex}
\end{align}
Eqs.~(\ref{eq:e_ap}) and (\ref{eq:e_inc}) are used for $ \eta_\mathrm{ sp, en }$.
Since the propagation between the telescope aperture and the entrance pupil is described with geometrical optics in this paper, $ \eta_\mathrm{ sp,en} $ might be considerably different from that calculated here due to diffraction.
The textbook \cite{Goldsmith} gives the exit pupil spillover efficiency for a fundamental-mode Gaussian beam with blockage $ \varepsilon $ in Eq.~(\ref{eq:sp_ex}).

\subsection{Evaluating at the entrance pupil} \label{sec3.1}
The beam coupling efficiency evaluated at the entrance pupil is written as
\begin{align}
\eta_\mathrm{ bcp } \left( \boldsymbol{ p }; \boldsymbol{ r }_\mathrm{ bw } \right) & = \frac{
\left| \int_\varepsilon^1 \mathrm{ d } \rho \int_0^{ 2 \pi } \mathrm{ d } \psi \rho E_\mathrm{ en } \left( \boldsymbol{ p };
\boldsymbol{ \rho } \right) E_\mathrm{ det }^{ \prime \ast } \left( \boldsymbol{ \rho }; \boldsymbol{ r }_\mathrm{ bw } \right) \right|^2 }
{ \left( \int_\varepsilon^1 \mathrm{ d } \rho \int_0^{ 2 \pi } \mathrm{ d } \psi \rho \left| E_\mathrm{ en }
\left( \boldsymbol{ p }; \boldsymbol{ \rho } \right) \right|^2 \right)
\left( \int_\varepsilon^1 \mathrm{ d } \rho \int_0^{ 2 \pi } \mathrm{ d } \psi \rho \left| E_\mathrm{ det }^\prime \left( \boldsymbol{ \rho };
\boldsymbol{ r }_\mathrm{ bw } \right) \right|^2 \right) }, \label{eq:def_coup}
\end{align}
where the electric field $ E_\mathrm{ det }^\prime \left( \boldsymbol{ \rho }; \boldsymbol{ r }_\mathrm{ bw }, w_\mathrm{ bw } \right) $ is the field on the entrance pupil originating from the feed.
The beam coupling efficiency is the most important quantity because of the close relation to a beam pattern.
Let us introduce the beam pattern as a function of direction cosines $ l $ and $ m $, $ \tilde{ P } ( l, m; \boldsymbol{ r }_\mathrm{ bw } ) $ (cf. \cite{Thompson1986}).
The numerator in Eq.~(\ref{eq:def_coup}) is proportional to $\tilde{P}(l,m)$ because the exponential function is equivalent to the incident field $ E_\mathrm{ en } ( \boldsymbol{ p } ) $.
As a result, we can relate the beam coupling efficiency with the beam pattern as follows:
\begin{align}
\begin{split}
\eta_\mathrm{ bcp } ( \boldsymbol{ p }; \boldsymbol{ r }_\mathrm{ bw } ) & = \frac{ \lambda^2 \tilde{ P }_n ( l, m; \boldsymbol{ r }_\mathrm{ bw } ) } { A_\mathrm{ en } \varOmega_\mathrm{ A } ( \boldsymbol{ r }_\mathrm{ bw } ) }, \\
A_\mathrm{ en } = \pi { R_\mathrm{ en } }^2 \left( 1 - \varepsilon^2 \right), ~
\tilde{ P }_\mathrm{ n } ( l, m; \boldsymbol{ r }_\mathrm{ bw } ) & = \frac{ \tilde{ P } ( l, m; \boldsymbol{ r }_\mathrm{ bw } ) }
{ \tilde{ P } ( l_0, m_0; \boldsymbol{ r }_\mathrm{ bw } ) }, ~
\varOmega_\mathrm{ A } = \iint \tilde{ P }_\mathrm{ n }
( l, m; \boldsymbol{ r }_\mathrm{ bw } ) \mathrm{ d } l \mathrm{ d } m,
\end{split}
\label{eq:eta_coup}
\end{align}
where $ A_\mathrm{ en } $, $ \tilde{ P }_\mathrm{ n } $, and $ \varOmega_\mathrm{ A } $ are the area of the entrance pupil, a normalized beam pattern, and a beam solid angle, respectively.
The direction cosines $ ( l_0, m_0 ) $ denotes the direction that makes $ \tilde{ P } ( l, m ) $ maximized.
When $ \tilde{ P }_n = 1 $, i.e., $ \boldsymbol{ p }_0 = \left( l_0, m_0 \right) $, Eq.~(\ref{eq:eta_coup}) reduces to a simple form,
\begin{align}
\eta_\mathrm{ A } ( \boldsymbol{ p }_0; \boldsymbol{ r }_\mathrm{ bw } ) = \eta_\mathrm{ sp, en } ( \boldsymbol{ p }_0 )
\eta_\mathrm{ sp, ex } (\boldsymbol{ r }_\mathrm{ bw } ) \frac{ \lambda^2 } { A_\mathrm{ p, en }
\varOmega_\mathrm{ A } ( \boldsymbol{ r }_\mathrm{ bw } ) }. \label{eq:eta_A}
\end{align}


\subsection{Evaluating at the exit pupil} \label{sec3.2}
We make an analytical expression of the aperture efficiency with the coefficients in Appendix~\ref{app2}.
By using Eqs.~(\ref{eq:e_ex_z}) and (\ref{eq:e_det_z}), the coupling efficiency is written as
\begin{align}
\eta_\mathrm{ bcp } \left( \boldsymbol{ p }; \boldsymbol{ r }_\mathrm{ bw } \right) & = \frac{ \left| \sum_{ m,n } { B_n }^m \left( \boldsymbol{ p }; \boldsymbol{ r }_\mathrm{ ref }; \varepsilon \right)
{ C_n^\ast }^m \left( \boldsymbol{ r }_\mathrm{ ref }; \boldsymbol{ r }_\mathrm{ bw }; w_\mathrm{ bw }; \varepsilon \right) \right|^2 }
{ \sum_{ n,m } \left| { B_n }^m \left( \boldsymbol{ p }; \boldsymbol{ r }_\mathrm{ ref }; \varepsilon \right) \right|^2 \sum_{ n, m }
\left| { C_n }^m \left( \boldsymbol{ r }_\mathrm{ ref }; \boldsymbol{ r }_\mathrm{ bw }; w_\mathrm{ bw }; \varepsilon \right) \right|^2 }.
\label{eq:bcp_at_exit}
\end{align}
The coefficients $ { B_n }^m $ hold information about the wavefront distorted by aberrations.
In terms of the denominator, Eqs.~(\ref{eq:e_ex_ref}) and (\ref{eq:e_ex_z}) yield
\begin{align}
\sum_{ n, m } \left| { B_n }^m \left( \boldsymbol{ p }; \boldsymbol{ r }_\mathrm{ ref }; \varepsilon \right) \right|^2 = 1 + O \left( W^3 \right).
\label{eq:B-sum}
\end{align}
Since we are considering the fundamental-mode Gaussian beam, the following is derived:
\begin{align}
\sum_{ p, q } \left| { C_p }^q \left( \boldsymbol{ r }_\mathrm{ bw } \right) \right|^2 = \frac{ e^{ - T_\mathrm{ e } \varepsilon^2 }
- e^{ - T_\mathrm{ e } } } { \pi \left( 1 - \varepsilon^2 \right) } + O \left( \sin^2 \theta_\mathrm{ bw } \right). \label{eq:C-sum}
\end{align}

The fundamental-mode Gaussian beam propagated obliquely through the system is expressed with Eqs.~(\ref{eq:e_det_z}) and (\ref{eq:Cpq}) with $ { D_0 }^0 = 1 $ for $ p^\prime = q^\prime = 0 $, and $ { D_{ p^\prime } }^{ q^\prime } = 0 $ for the others.
The coefficients $ { C_p }^q $ for the fundamental mode are obtained,
\begin{align}
\begin{array}{ll}
{ C_p }^0 = F_0 \sqrt{ p + 1 } \left[ \tilde{ R }_{ p } {}^0 \left( I_{ p } \right) + F_2 \tilde{ R }_{ p } {}^0 \left( I_{ p + 2 } \right) \right], & ( p : \mathrm{ even } ) \\
{ C_p }^{ \pm 1 } = F_0 \sqrt{ p + 1 } \left[ F_1^\mp \tilde{ R }_{ p } {}^1 \left( I_{ p + 1 } \right)
+ F_3^\mp \tilde{ R }_{ p } {}^1 \left( I_{ p + 3 } \right) \right], & ( p : \mathrm{ odd } ) \\
{ C_p }^{ | q | } = 0. & ( | q | \geq 2 )
\end{array}
\label{eq:Cpq_fundamental_mode}
\end{align}
The coefficient $ F_2 $ corresponds to defocus as shown in Eq.~(\ref{eq:parameters}). We therefore assume compensation by the longitudinal adjustment of a feed such that $ F_2 = 0 $.
Factoring out $ { C_0^\ast }^0 $, which is calculated with Eq.~(\ref{eq:Cpq_fundamental_mode}), from the coupling efficiency and using Eqs.~(\ref{eq:sp_en}), (\ref{eq:sp_ex}), (\ref{eq:bcp_at_exit}) to (\ref{eq:C-sum}), (\ref{eq:Bpq-1}), and (\ref{eq:Bpq-2}), we obtain the analytical expression of the aperture efficiency affected by the Seidel aberrations,
\begin{align}
& \eta_\mathrm{ A } = \frac{ 4 { R_\mathrm{ en } }^2 \left( e^{ - \frac{ T_\mathrm{ e } } { 2 } \varepsilon^2 }
- e^{ - \frac{ T_\mathrm{ e } } { 2 } } \right)^2 } { { R_\mathrm{ ap } }^2 T_\mathrm{ e } } \cos \varTheta
\left| 1 + \frac{ j k } { { C_0^\ast }^0 } \left( { A_1 }^{ 1 } { C^\ast_1 }^{ 1 } + { A_1 }^{ -1 } { C^\ast_1 }^{ -1 }
+ { A_2 }^{ 0 } { C^\ast_2 }^{ 0 } + { A_3 }^{ 1 } { C^\ast_3 }^{ 1 } \right. \right. \nonumber \\
& \left. \left. \hspace{ 45 mm } + { A_3 }^{ -1 } { C^\ast_3 }^{ -1 }
+ { A_4 }^{ 0 } { C^\ast_4 }^{ 0 } \right) - \frac{ k^2 } { { C_0^\ast }^0 } \left\{ G^0 + \sum_{ s = \pm 1 } G^s \right\} \right|^2, \label{eq:ap_eff}
\end{align}
where
\begin{align}
& G^0 \coloneq \left( \frac{ { C_0^\ast }^0 } { 2 } + \frac{ { C_4^\ast }^0 } { \sqrt{ 5 } } \right) \left( { A_2 }^0 \right)^2
+ \left( \frac{ 2 } { \sqrt{ 5 } } { C_2^\ast }^0 + \frac{ 3 \sqrt{ 3 } } { \sqrt{ 35 } } { C_6^\ast }^0 \right) { A_2 }^0 { A_4 }^0
+ \left( \frac{ { C_0^\ast }^0 } { 2 } + \frac{ \sqrt{ 5 } } { 7 } { C_4^\ast }^0 + \frac{ 3 }{ 7 } { C_8^\ast }^0 \right) \left( { A_4 }^0 \right)^2 \nonumber \\
& + \left( { C_0^\ast }^0 + \frac{ [ 1 , - 1 ] } { \sqrt{ 3 } [ 1 , 1 ] } { C_2^\ast }^0 \right) { A_1 }^1 { A_1 }^{ -1 }
+ \left( { C_0^\ast }^0 + \frac{ \sqrt{ 3 }[ 1 , 1 ] [ 1 , - 1 ] } { 2 [ 1 , 1 , 1 ] } { C_2^\ast }^0
+ \frac{ [ 1 , - 1 ]^2 } { 2 \sqrt{ 5 } [ 1 , 1 , 1 ] } { C_4^\ast }^0 \right) { A_2 }^2 { A_2 }^{ -2 } \nonumber \\
& + \left( { C_0^\ast }^0 + \frac{ [ 1 , - 1 ] [ 1 , - 8 , 1 ] } { 5 \sqrt{ 3 } [ 1 , 1 ] [ 1 , 4 , 1 ] } { C_2^\ast }^0
+ \frac{ [ 1, 10, 1 ] } { \sqrt{ 5 } [ 1, 4, 1 ] } { C_4^\ast }^0 + \frac{ 9 [ 1, 1 ] [ 1, -1 ] } { 5 \sqrt{ 7 } [ 1, 4, 1 ] } { C_6^\ast }^0 \right)
{ A_3 }^1 { A_3 }^{ -1 }, \label{eq:G} \\
& G^s \coloneq \frac{ \left( [ 1 , - 1 ] { C_1^\ast }^s + \sqrt{ 2 [ 1 , 4 , 1 ] } { C_3^\ast }^s \right) { A_1 }^s { A_2 }^{ 0 } } { \sqrt{ 3 } [ 1 , 1 ] }
+ \frac{ \left( \sqrt{ 2 } [ 1 , - 1 ] { C_3^\ast }^s + \sqrt{ 3 [ 1 , 8 , 1 ] } { C_5^\ast }^s \right) { A_1 }^{ s } { A_4 }^0 } { \sqrt{ 5 [ 1 , 4 , 1 ] } } \nonumber \\
& + \frac{ \left( 2 [ 1 , 1 , 1 ] { C_1^\ast }^s + [ 1 , - 1 ] \sqrt{ [ 1 , 4 , 1 ] } { C_3^\ast }^s \right) { A_1 }^{ -s } { A_2 }^{ 2 s } } { \sqrt{ 3 } [ 1 , 1 ] \sqrt{ [ 1 , 1 , 1 ] } }
+ \left( \frac{ [ 1 , 4 , 1 ] } { \sqrt{ 3 } [ 1 , 1 ] } { C_2^\ast }^0
+ \frac{ [ 1 , - 1 ] } { \sqrt{ 5 } } { C_4^\ast }^0 \right) \frac{ \sqrt{ 2 } { A_1 }^s { A_3 }^{ -s } } { \sqrt{ [ 1, 4 ,1 ] } } \nonumber \\
& + \left( \frac{ \sqrt{ 2 [ 1 , 4 , 1 ] } } { \sqrt{ 3 } [ 1 , 1 ] } { C_1^\ast }^s + \frac{ [ 1 , - 1 ]
[ 1 , - 8 , 1 ] } { 5 \sqrt{ 3 } [ 1 , 1 ] [ 1 , 4 , 1 ] } { C_3^\ast }^s + \frac{ 3 \sqrt{ 2 } [ 1 , 1 ] \sqrt{ [ 1 , 8 , 1 ] } }
{ 5 [ 1 , 4 , 1 ] } { C_5^\ast }^s \right) { A_2 }^0 { A_3 }^{ s } \label{eq:G2} \\
& + \left( \frac{ [ 1 , - 1 ] \sqrt{ [ 1 , 4 , 1 ] } }{ \sqrt{ 6 } [ 1 , 1 ] }
{ C_1^\ast }^s + \frac{ 4 [ 2 , 10 , 21 , 10 , 2 ] { C_3^\ast }^s } { 5 \sqrt{ 3 } [ 1 , 1 ] [ 1 , 4 , 1 ] }
+ \frac{ 3 [ 1 , 1 ] [ 1 , - 1 ] \sqrt{ [ 1 , 8 , 1 ] } } { 5 \sqrt{ 2 } [ 1 , 4 , 1 ] } { C_5^\ast }^s \right) \frac{ { A_2 }^{ 2 s } { A_3 }^{ -s } } { \sqrt{ [ 1 , 1 , 1 ] } } \nonumber \\
& + \left( [ 1 , - 1 ] { C_1^\ast }^s +
\frac{ [ 1 , 10 , 1 ] { C_3^\ast }^s } { \sqrt{ 2 [ 1 , 4 , 1 ] } }
+ \frac{ \sqrt{ 3 } [ 1 , - 1 ] [ 1 , - 10 , 1 ] { C_5^\ast }^s } { 7 \sqrt{ [ 1 , 4 , 1 ] [ 1 , 8 , 1 ] } }
+ \frac{ 9 \sqrt{ [ 1, 16, 36, 16, 1 ] } { C_7^\ast }^s } { 7 \sqrt{ [ 1, 8, 1 ] } } \right)
\frac{ \sqrt{ 2 } { A_3 }^{ s } { A_4 }^0 } { \sqrt{ 5 [ 1, 4, 1 ] } }. \nonumber
\end{align}
To save the space, we have introduced the following notation for the $ \varepsilon^2 $ polynomials and used it:
\begin{align}
& [ a, b, c, \cdots] \coloneq a + b \varepsilon^2 + c \varepsilon^4 + \cdots. \label{eq:eps_pol}
\end{align}
Eq.~(\ref{eq:Cpq_fundamental_mode}) gives $ { C_p }^q $ for the fundamental-mode Gaussian beam to Eqs.~(\ref{eq:ap_eff}) to (\ref{eq:G2}).
The products $ { A_1 }^{ \pm 1 } { C^\ast_1 }^{ \pm 1 } $ and $ { A_2 }^0 { C_2^\ast }^0 $ correspond to the effects of tip-tilt and defocus,
respectively, which are strongly dependent on $ \boldsymbol{ r }_\mathrm{ bw } $.
Let us focus on the first order $ { A_n }^m $.
If the beam waist of the feed is located such that the following conditions are satisfied,
\begin{align}
{ A_2 }^0 ( \boldsymbol{ p }; \boldsymbol{ r }_\mathrm{ ref }; \varepsilon ) & = - \frac{ { C^\ast_4 }^0 ( \boldsymbol{ r }_\mathrm{ ref };
\boldsymbol{ r }_\mathrm{ bw }; w_\mathrm{ bw }; \varepsilon ) } { { C^\ast_2 }^0 ( \boldsymbol{ r }_\mathrm{ ref }; \boldsymbol{ r }_\mathrm{ bw };
w_\mathrm{ bw }; \varepsilon ) } { A_4 }^0 ( \boldsymbol{ p }; \boldsymbol{ r }_\mathrm{ ref }; \varepsilon ), \label{eq:for_sph} \\
{ A_1 }^{ \pm 1 } ( \boldsymbol{ p }; \boldsymbol{ r }_\mathrm{ ref }; \varepsilon ) & = - \frac{ { C^\ast_3 }^{ \pm 1 }
( \boldsymbol{ r }_\mathrm{ ref }; \boldsymbol{ r }_\mathrm{ bw }; w_\mathrm{ bw }; \varepsilon ) } { { C^\ast_1 }^{ \pm 1 }
( \boldsymbol{ r }_\mathrm{ ref }; \boldsymbol{ r }_\mathrm{ bw }; w_\mathrm{ bw }; \varepsilon ) }
{ A_3 }^{ \pm 1 } ( \boldsymbol{ p }; \boldsymbol{ r }_\mathrm{ ref }; \varepsilon ), \label{eq:for_coma}
\end{align}
then, the first order terms vanish.
Eqs.~(\ref{eq:for_sph}) and (\ref{eq:for_coma}) represent the conditions for reducing the effect of spherical aberration and coma, respectively.

We can calculate the coefficients $ { C_p }^q $ for an arbitrary feed, though we have limited ourselves to the fundamental-mode Gaussian beam case.
If an asymmetric feed pattern is considered, e.g., a diagonal horn, we will obtain the conditions for reducing the effect of astigmatism, $ { A_2}^{ \pm 2 } $.


\section{Verification} \label{sec4}
Eq.~(\ref{eq:ap_eff}) and the conditions in Eqs.~(\ref{eq:for_sph}) and (\ref{eq:for_coma}) are verified with numerical simulations.
We compare the aperture efficiency evaluated using ray tracing \cite{Zemax} and the PO simulation \cite{GRASP8}.

\subsection{Model and calculation} \label{sec4.1}
\begin{figure}[!t]
\centering
\includegraphics[ width = 0.4 \hsize ]{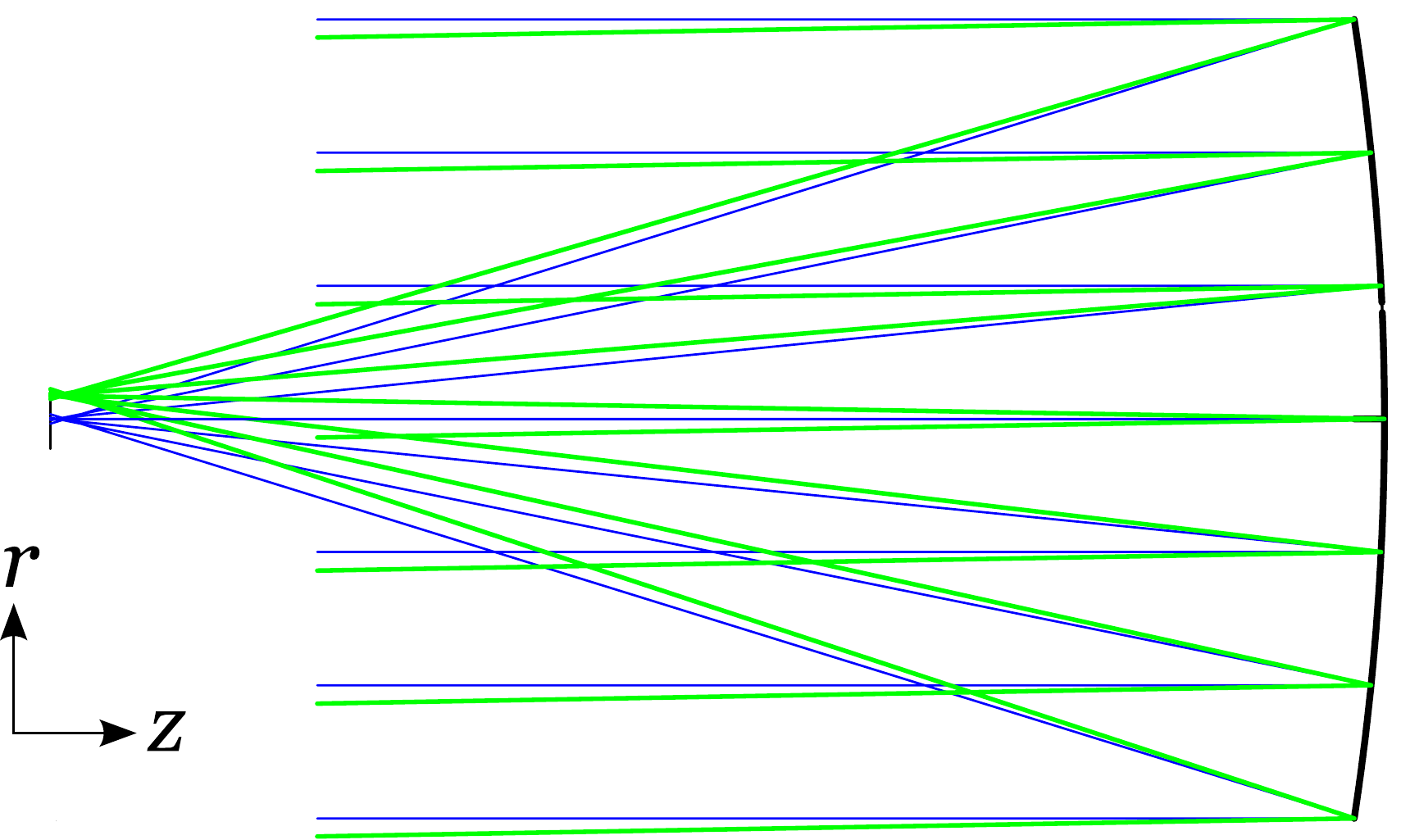}
\caption{Schematics of the model used for verification. The blue rays represent the light with $ \boldsymbol{ p } = (0, 0) $. The green ones are the light for $ \boldsymbol{ p } = ( \sin 1^\circ, 0) $.}
\label{fig2}
\end{figure}

We use a simple system composed of a spherical mirror with a circular aperture ($ \varepsilon = 0 $, Fig.~\ref{fig2}).
The radius of curvature of the mirror is $ -1000 $\,mm and its diameter is $ 300 $\,mm.
The wavelength in simulation is 200\,$\mu$m.
The incident direction are $ \boldsymbol{ p } = ( 0, 0 ) $ and $ ( \sin 1^\circ, 0 ) $.
The beam waist position $ \boldsymbol{ r }_\mathrm{ bw } $ is determined such that $ R ( 0; z_\mathrm{ bw } ) = z_\mathrm{ ref } $, $ \theta_\mathrm{ bw } = \theta_\mathrm{ ref } $, and $ \phi_\mathrm{ bw } = \phi_\mathrm{ ref } = 0 $.
The reference sphere center $ \boldsymbol{ r }_\mathrm{ ref } $ is located at the points where Eqs.~(\ref{eq:for_sph}) and (\ref{eq:for_coma}) hold for edge taper of 15\,dB (Case~1), where the Strehl ratio without apodization is maximized (Case~2), and where the Gaussian image point is located (Case~3).
In the PO simulation, a fundamental-mode Gaussian beam is used.
The edge taper is approximately set to 5, 10, 15, and 20\,dB for every case.
The exact values of the spillover efficiency, edge taper, and beam solid angles are derived from the PO simulation.
Then, the aperture efficiency is calculated using Eq.~(\ref{eq:eta_A}).

The aberration coefficients ${ A_n }^m $ are derived from ray tracing simulation.
The aperture efficiency is calculated using Eq.~(\ref{eq:ap_eff}) as a function of ${ A_n }^m $ and the parameter $ T_\mathrm{ e } $ which is defined in Eq.~(\ref{eq:Te}) and corresponds to the edge taper.
The Strehl ratio is calculated with ${ A_n }^m $ \cite{Mahajan1981, Mahajan1983},
$ S \simeq \exp \left[ - k^2 { W_\mathrm{ dev } }^2 ( \boldsymbol{ p }; \boldsymbol{ r }_\mathrm{ ref } ) \right] $, where $ W_\mathrm{ dev } $ is the deviation of the wavefront aberrations across the exit pupil and given by $ { W_\mathrm{ dev } }^2 \left( \boldsymbol{ p }; \boldsymbol{ r }_\mathrm{ ref } \right) = \sum_{m,n} \left| { A_n }^m ( \boldsymbol{ p }; \boldsymbol{ r }_\mathrm{ ref }; \varepsilon ) \right|^2 $.

\begin{table}[!t]
\centering
\caption{Aperture efficiency obtained with Eqs.~(\ref{eq:ap_eff}) and (\ref{eq:eta_A}) for $ \boldsymbol{ p } = (0,0)$.}
\label{tab2}
\begin{tabular}{c|c|c|c|c|c|c} \hline
case & $ ( r_\mathrm{ bw }, z_\mathrm{ bw } )$ in mm & Strehl ratio & Edge taper & $ \eta_\mathrm{ A, an } $ & $ \eta_\mathrm{ A, PO } $ &
$ \frac{ \eta_\mathrm{ A, an } } { \eta_\mathrm{ A, PO } } - 1 $ \\ \hline
1 &$(0,-497.329)$ & 0.9114 & 5.463\,dB & 0.6339 & 0.6380 & $-0.6$\% \\
& & & 10.356\,dB & 0.7467 & 0.7475 & $-0.1$\% \\
& & & 15.246\,dB & 0.7168 & 0.7149 & $0.3$\% \\
& & & 20.135\,dB & 0.6454 & 0.6421 & $0.5$\% \\ \hline
2 & $(0,-497.168)$ & 0.9150 & 5.464\,dB & 0.6324 & 0.6363 & $-0.6$\% \\
& & & 10.356\,dB & 0.7409 & 0.7417 & $-0.1$\% \\
& & & 15.246\,dB & 0.7080 & 0.7062 & $0.3$\% \\
& & & 20.135\,dB & 0.6353 & 0.6321 & $0.5$\% \\ \hline
3 & $(0,-500.000)$ & 0.2527 & 5.460\,dB & 0.1162 & 0.2057 & $-43.5$\% \\
& & & 10.354\,dB & 0.2285 & 0.3013 & $-24.1$\% \\
& & & 15.247\,dB & 0.3299 & 0.3509 & $-6.0$\% \\
& & & 20.139\,dB & 0.4051 & 0.3720 & $8.9$\% \\ \hline
\end{tabular}
\end{table}

\subsection{Results} \label{sec4.2}

\begin{figure}[!t]
\centering
\includegraphics[width=\hsize]{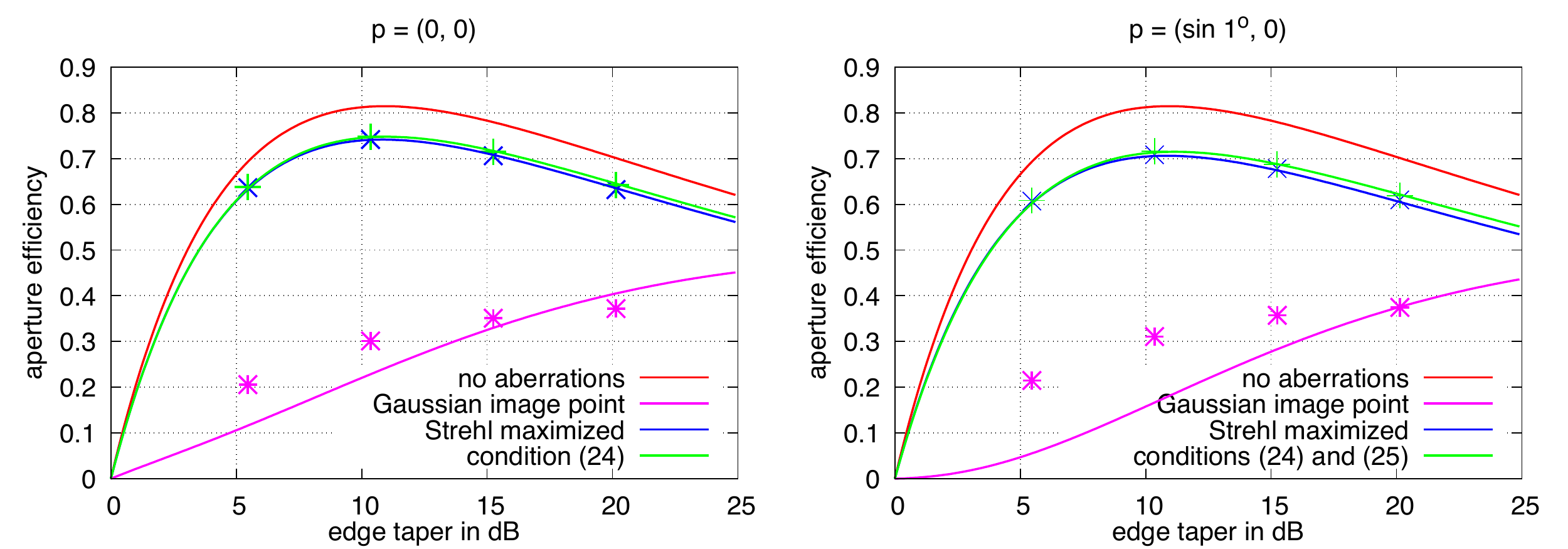} \\
\caption{Aperture efficiency predicted by Eq.~(\ref{eq:ap_eff}) (curves) and the PO simulation (dots). The red line represents no aberrations, green Case~1, blue Case~2, and magenta Case~3.}
\label{fig3}
\end{figure}

Tables~\ref{tab2} and \ref{tab3} are for $ \boldsymbol{ p } = (0, 0) $ and $ ( \sin 1^\circ, 0) $, respectively.
Both tables contain the beam waist position $ \boldsymbol{ r }_\mathrm{ bw }$, Strehl ratio, edge taper, aperture efficiency from ray tracing, $ \eta_\mathrm{ A, an } $, aperture efficiency from the PO simulations, $ \eta_\mathrm{ A, PO } $, and difference $ \eta_\mathrm{ A, an }/ \eta_\mathrm{ A, PO } - 1 $.
The aperture efficiencies estimated using Eq.~(\ref{eq:ap_eff}) agree with those calculated from the PO simulation for the higher Strehl ratios.
Fig.~\ref{fig3} shows the points obtained from the PO simulation and the theoretical curves predicted by Eq.~(\ref{eq:ap_eff}) as a function of the edge taper for both incident angles.
The red lines represent the aperture efficiency without any aberrations for reference.
The green lines (case 1) give the highest aperture efficiency in the cases considered here with the same values of $ { A_n }^m $.
Note that the difference between cases 1 and 2 was small but case 1 provided the higher aperture efficiency.
That is, the optimization of the feed position in terms of the Strehl ratio does not necessarily maximize the aperture efficiency.

\begin{figure}[!t]
\centering
\includegraphics[ width = 0.45 \hsize ]{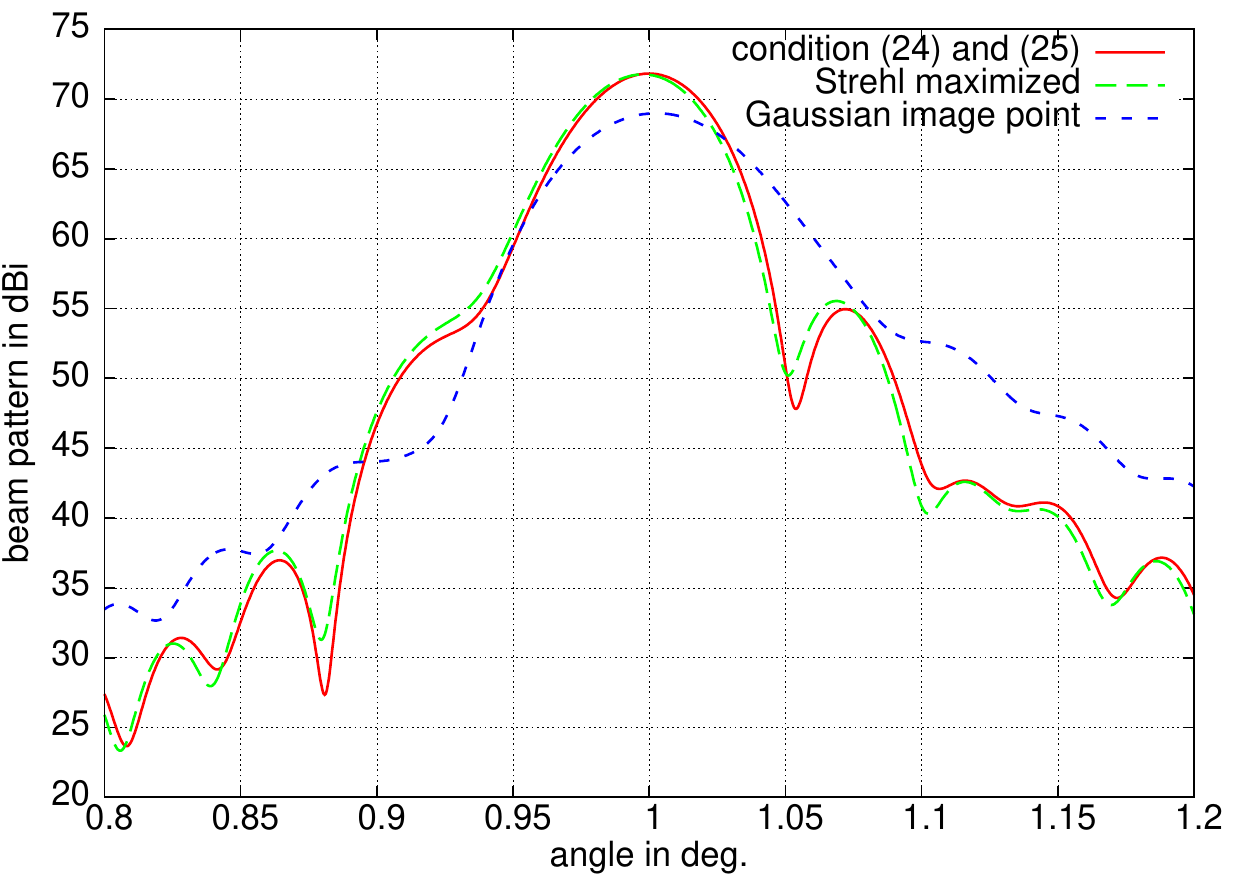}
\caption{Beam patterns for $ \boldsymbol{ p } = ( \sin 1^\circ, 0 )$.
The peak shifts from $ \varTheta = 1^\circ $ are $-0^{\prime\prime}.50$, $-8^{\prime\prime}.39$, and $9^{\prime\prime}.47$ for cases 1 (red), 2 (green), and 3 (blue), respectively. }
\label{fig4}
\end{figure}

Fig.~\ref{fig4} shows the beam patterns on the meridional plane for $ \boldsymbol{ p } = ( \sin 1^\circ, 0) $.
The peak gains and positions are different among the cases~1, 2, and 3.
When the condition in Eq.~(\ref{eq:for_coma}) holds, Fig.~\ref{fig4} indicates that we can reduce pointing errors due to the third-order coma.
The condition in Eq.~(\ref{eq:for_coma}) is practically useful to design a wide FOV radio telescope.

\begin{table}[!t]
\centering
\caption{Aperture efficiency obtained using Eqs.~(\ref{eq:ap_eff}) and (\ref{eq:eta_A}) for $ \boldsymbol{p} =( \sin 1^\circ,0)$.}
\label{tab3}
\begin{tabular}{c|c|c|c|c|c|c} \hline
case & $ ( r_\mathrm{ bw }, z_\mathrm{ bw } ) $ in mm & Strehl ratio & Edge taper & $ \eta_\mathrm{ A, an } $ & $ \eta_\mathrm{ A, PO } $ &
$ \frac{ \eta_\mathrm{ A, an } } { \eta_\mathrm{ A, PO } } - 1 $ \\ \hline 
1 & $(8.789,-497.256)$ & 0.8615 & 5.463\,dB & 0.6020 & 0.6082 & $-1.0$\% \\
& & & 10.354\,dB & 0.7133 & 0.7161 & $-0.4$\% \\
& & & 15.243\,dB & 0.6879 & 0.6875 & $0.1$\% \\
& & & 20.131\,dB & 0.6216 & 0.6193 & $0.4$\% \\ \hline
2 & $(8.806,-497.095)$ & 0.8727 & 5.463\,dB & 0.6019 & 0.6070 & $-0.8$\% \\
& & & 10.354\,dB & 0.7055 & 0.7083 & $-0.4$\% \\
& & & 15.243\,dB & 0.6743 & 0.6781 & $-0.6$\% \\
& & & 20.131\,dB & 0.6051 & 0.6096 & $-0.7$\% \\ \hline
3 & $(8.728,-500.000)$ & 0.1571 & 5.459\,dB & 0.0553 & 0.2149 & $-74.3$\% \\
& & & 10.353\,dB & 0.1665 & 0.3107 & $-46.4$\% \\
& & & 15.244\,dB & 0.2831 & 0.3573 & $-20.8$\% \\
& & & 20.135\,dB & 0.3757 & 0.3747 & $0.3$\% \\ \hline
\end{tabular}
\end{table}

\begin{table}[!b]
\centering
\caption{Relative magnitude of the third-order term with respect to the zeroth-order term and corresponding Strehl ratio.}
\label{tab5}
\begin{tabular}{c|c|c} \hline
$ W_\mathrm{ dev }$ & $ | -jk^3 { W_\mathrm{ dev } }^3 / 6 | $ & Strehl ratio \\ \hline
$ \lambda / 20 $ & 0.5\% & 0.906 \\
$ \lambda / 16.5 $ & 0.9\% & 0.865 \\
$ \lambda / 13.3 $ & 1.8\% & 0.800 \\
$ \lambda / 5 $ & 33.1\% & 0.204 \\ \hline
\end{tabular}
\end{table}

\section{Discussion} \label{sec5}
\subsection{Approximation precision and the Strehl ratio} \label{sec5.1}
In this subsection, we address how precise the analytical expression works and how the Strehl ratio relates to it.

The results in Section \ref{sec4.2} indicate that the aperture efficiency estimated with Eq.~(\ref{eq:ap_eff}) agrees with that calculated from the PO simulation.
Eq.~(\ref{eq:ap_eff}) has been derived under the approximations:
only the Seidel aberrations are taken into account and the Taylor series is terminated at the second order.
The higher order aberrations are quite small in our verification.
Therefore, let us focus on the order of the Taylor expansion in Eq.~(\ref{eq:e_ex_z}).
The omitted terms were the third or higher orders of $ W $.
The absolute value of the largest term $ - j k^3 W^3 / 6 $ would be estimated by replacement of $ W $ with the standard deviation of wavefront errors, $ W_\mathrm{ dev } $.
Table~\ref{tab5} shows the deviation of the wavefront error, the estimated third order from $ W_\mathrm{ dev } $, and the corresponding Strehl ratio.
The discrepancy $ \eta_\mathrm{ A, an } / \eta_\mathrm{ A, PO } - 1 $ in Tables~\ref{tab2} and \ref{tab3} seem close to the magnitude of the third-order term in Table \ref{tab5} for higher Strehl ratios.
The Strehl ratio implies the precision of Eq.~(\ref{eq:ap_eff}).
If the Strehl ratio is 0.8, we will be able to estimate the aperture efficiency with a precision of 2\% or so.


\subsection{Applications} \label{sec5.2}
We can extract various information on an optical system by selecting a proper parameter as a free parameter.
We briefly look into potential applications in this subsection.

When an incident direction $ \boldsymbol{ p } $ is a free parameter and the other parameters are fixed,
we can obtain the beam pattern as shown in Eq.~(\ref{eq:eta_coup}).
All we have to do is to calculate $ { A_n }^m ( \boldsymbol{ p }; \boldsymbol{ r }_\mathrm{ ref }; \varepsilon ) $ as a function of $ \boldsymbol{ p } $ using ray tracing software. The higher orders of $ { A_n }^m $ may be added if need be.

Let us focus on the focal plane and consider the case when aperture efficiency is a function of the beam waist position $ \boldsymbol{ r }_\mathrm{ bw } $, which can be regarded as a detector position, and the other parameters are fixed.
The detector position determines the coefficients $ { C_p }^q $.
The dependence of aperture efficiency on $ \boldsymbol{ r }_\mathrm{ bw } $ allows us to estimate the tolerance of the detector position.
In a special case, when a detector has an isotropic sensitivity we will obtain a point spread function.

The aberration coefficients $ { A_n }^m $ can free parameters with $ \boldsymbol{ p } $ and $ \boldsymbol{ r }_\mathrm{ bw } $ fixed.
This situation happens when the optical elements are misaligned and deformed.
In that case, we can apply Eq.~(\ref{eq:ap_eff}) to tolerance analysis with ray tracing.
Generally, tolerance analysis requires numerous cases of misalignment and deformation, and therefore, it is unreasonable to use full-wave simulation, which consumes considerable amount of computing resources.
Eq.~(\ref{eq:ap_eff}) can give the aperture efficiencies without full-wave simulation.

Finally, the limitation of this analytical expression is addressed.
The assumption used in this study are that the propagation from the telescope aperture to the exit pupils can be described with geometrical optics and the feed beam is described with Gaussian beam theory, which is an equivalent approximation to the Fresnel diffraction theory.
Therefore, we need full-wave simulation in the following cases: diffraction effects at the edges of optical elements are significant,
a higher order approximation of diffraction is needed compared to the Fresnel diffraction theory, and polarization has to be evaluated.


\section{Conclusion} \label{sec6}
Aperture efficiency is one of the figures of merit of a radio telescope.
We explicitly show that it depends on the incident direction $ \boldsymbol{ p } $, the position of detectors $ \boldsymbol{ r }_\mathrm{ bw } $, and the feed pattern.
The wavefront errors and feed pattern are expanded into a series of the Zernike annular polynomials, whose coefficients are given as a function of either  $ \boldsymbol{ p } $ and $ \boldsymbol{ r }_\mathrm{ bw } $, respectively.
The expansion enables us to derive the analytical expression of the aperture efficiency affected by the Seidel aberrations.
If the Strehl ratio without apodization is greater than 0.8, this expression gives aperture efficiency with a precision of 2\% from ray tracing simulation.
In addition, we derive the useful conditions required to reduce the effects of spherical aberration and coma.
In particular, the condition for reducing coma avoids the pointing error caused by coma.
If a non-axially symmetric feed pattern is assumed, a condition to reduce astigmatism may be derived.
The expression can be applied for the evaluation of a beam pattern and tolerance analysis.

\appendix
\section{Zernike annular polynomials} \label{app1}
The Zernike annular polynomials \cite{Mahajan1981} are of the form
\begin{align}
& {Z_n}^m(\rho,\psi;\varepsilon) = \sqrt{ n + 1 } { R_n }^{ | m | } \left( \rho, \varepsilon \right) \exp \left( j m \psi \right),
\label{eq:zap}
\end{align}
where $ m $ and $ n $ are integers such that $ n \geq | m | $ and $ n - | m | $ is even.
The domain is $ 0 \leq \varepsilon \leq \rho \leq 1 ~ ( \varepsilon < 1 ) $ and $ 0 \leq \theta \leq 2 \pi $.
The normalization is given by
\begin{align}
\int_\varepsilon^1 \mathrm{ d } \rho \int_0^{ 2 \pi } \mathrm{ d } \psi \, \rho { Z_n }^m( \rho, \psi; \varepsilon ) { Z_p }^{ q \ast }
( \rho, \psi; \varepsilon ) = \pi \left( 1 - \varepsilon^2 \right) \delta_{ np } \delta_{ mq }.
\label{eq:zap_norm}
\end{align}
The polynomials which were not demonstrated in \cite{Mahajan1981} are listed in Eq.~(\ref{eq:Rpq}).
The notation for the $\varepsilon^2$ polynomials in Eq.~(\ref{eq:eps_pol}) is used.
\begin{align}
{ R_5 }^1 ( \rho; \varepsilon ) & = \cfrac{ 10 [ 1, 4, 1 ] \rho^5 - 12 [ 1, 4, 4, 1 ] \rho^3 + 3 [ 1, 4, 10, 4, 1 ] \rho }
{ [ 1, - 1 ]^2 \sqrt{ [ 1, 4, 1 ] [ 1, 9, 9, 1 ] } }, \nonumber \\
{ R_5 }^3 ( \rho; \varepsilon ) & = \cfrac{ 5 [ 1, 1, 1, 1 ] \rho^5 - 4 [ 1, 1, 1, 1, 1 ] \rho^3 } { [ 1, - 1 ] \sqrt{ [ 1, 1, 1, 1 ]
[ 1, 4, 10, 20, 10, 4, 1 ] } }, \nonumber \\
{ R_6 }^2 ( \rho; \varepsilon ) & = \cfrac{ 15 [ 1, 4, 10, 4, 1 ] \rho^6 - 20 [ 1, 4, 10, 10, 4, 1 ] \rho^4 + 6 [ 1, 4, 10, 20, 10, 4, 1 ]
\rho^2 } { [ 1, - 1 ]^2 \sqrt{ [ 1, 4, 10, 4, 1 ] [ 1, 9, 45, 65, 45, 9, 1 ] } }, \label{eq:Rpq} \\
{ R_7 }^1 ( \rho; \varepsilon ) & = \cfrac{ 35 [ 1, 9, 9, 1 ] \rho^7 - 60 [ 1, 9, 15, 9, 1 ] \rho^5 +30 [ 1, 9, 25, 25, 9, 1 ] \rho^3
- 4 [ 1, 9, 45, 65, 45, 9, 1 ] \rho } { [ 1, - 1 ]^3 \sqrt{ [ 1, 9, 9, 1 ] [ 1, 16, 36, 16, 1 ] } }, \nonumber \\
{ R_8 }^0 ( \rho; \varepsilon ) & = \cfrac{ 70 \rho^8 - 140 [ 1, 1 ] \rho^6 +90 [ 1, 8/3, 1 ] \rho^4 - 20 [ 1, 6, 6, 1 ] \rho^2
+ [ 1, 16, 36, 16, 1 ] } { [ 1, - 1 ]^4 }. \nonumber
\end{align}

\section{Coefficients $ { B_n }^m $ for the Seidel aberrations} \label{app2}
The notation for the $\varepsilon$ polynomials in Eq.~(\ref{eq:eps_pol}) is used.
\begin{align}
{ B_0 }^0 & = 1 - \frac{ k^2 } { 2 } \left[ \left( { A_2 }^0 \right)^2 + \left( { A_4 }^0 \right)^2 + 2 { A_1 }^1 { A_1 }^{ -1 }
+ 2 { A_2 }^2 { A_2 }^{ -2 } + 2 { A_3 }^1 { A_3 }^{ -1 } \right], \nonumber \\
{ B_1 }^{ \pm 1 } & = j k { A_1 }^{ \pm 1 } - k^2 \left( \frac{ [ 1 , - 1 ] } { \sqrt{ 3 } [ 1 , 1 ] } { A_1 }^{ \pm 1 } { A_2 }^0
+ \frac{ 2 \sqrt{ [ 1 , 1 , 1 ] } } { \sqrt{ 3 } [ 1 , 1 ] } { A_1 }^{ \mp 1 } { A_2 }^{ \pm 2 } \right. \nonumber \\
&\hspace{ 3mm } \left. + \frac{ \sqrt{ 2 [ 1 , 4 , 1 ] } } { \sqrt{ 3 } [ 1 , 1 ] }{ A_2 }^0 { A_3 }^{ \pm 1 } + \frac{ [ 1 , - 1 ]
\sqrt{ [ 1 , 4 , 1 ] } }{ \sqrt{ 6 } [ 1 , 1 ] \sqrt{ [ 1 , 1 , 1 ] } } { A_2 }^{ \pm 2 } { A_3 }^{ \mp 1 }
+ \frac{ \sqrt{ 2 }[ 1 , - 1 ] }{ \sqrt{ 5 [ 1 , 4 , 1 ] } }{ A_3 }^{ \pm 1 } { A_4 }^0 \right), \label{eq:Bpq-1} \\
{ B_2 }^0 & = j k { A_2 }^0 - k^2 \left[ \frac{ [ 1 , - 1 ] } { \sqrt{ 3 } [ 1 , 1 ] } { A_1 }^1 { A_1 }^{ -1 }
+ \frac{ \sqrt{ 2 [ 1 , 4 , 1 ] } } { \sqrt{ 3 } [ 1 , 1 ] } \left( { A_1 }^1 { A_3 }^{ -1 } + { A_1 }^{ -1 } { A_3 }^{ 1 } \right) \right. \nonumber \\
& \hspace{ 3mm } \left. 
+ \frac{ 2 } { \sqrt{ 5 } } { A_2 }^0 { A_4 }^0 + \frac{ \sqrt{ 3 }[ 1 , 1 ] [ 1 , - 1 ] } { 2 [ 1 , 1 , 1 ] } { A_2 }^2 { A_2 }^{ -2 }
+ \frac{ [ 1 , - 1 ] [ 1 , - 8 , 1 ] } { 5 \sqrt{ 3 } [ 1 , 1 ] [ 1 , 4 , 1 ] } { A_3 }^1 { A_3 }^{ -1 } \right], \nonumber
\end{align}
\begin{align}
{ B_2 }^{ \pm 2 } & = j k { A_2 }^{ \pm 2 } - k^2 \left[
\frac{ \sqrt{ [ 1 , 1 , 1 ] } } { \sqrt{ 3 } [ 1 , 1 ] } \left( { A_1 }^{ \pm 1 } \right)^2
+ \frac{ [ 1 , - 1 ] \sqrt{ [ 1 , 4 , 1 ] } } { \sqrt{ 6 } [ 1 , 1 ] \sqrt{ [ 1 , 1 , 1 ] } } { A_1 }^{ \pm 1 } { A_3 }^{ \pm 1 } \right. \nonumber \\
&\hspace{ 3mm } \left. + \frac{ \sqrt{ 3 } [ 1 , 1 ] [ 1 , - 1 ] } { 2 [ 1 , 1 , 1 ] } { A_2 }^0 { A_2 }^{ \pm 2 } + \frac{ [ 1 , - 1]^2 { A_2 }^{ \pm 2 } { A_4 }^0 } { 2 \sqrt{ 5 } [ 1 , 1 , 1 ] } + \frac{ 2 [ 2 , 10 , 21 , 10 , 2 ] \left( { A_3 }^{ \pm 1 } \right)^2 } { 5 \sqrt{ 3 } [ 1 , 1 ] [ 1 , 4 , 1 ] \sqrt{ [ 1 , 1 , 1 ] } } \right], \nonumber \\
{ B_3 }^{ \pm 1 } & = j k { A_3 }^{ \pm 1 } - k^2 \left(
\frac{ \sqrt{ 2 [ 1 , 4 , 1 ] } }{ \sqrt{ 3 } [ 1 , 1 ] } { A_1 }^{ \pm 1 } { A_2 }^0
+ \frac{ [ 1 , - 1 ] \sqrt{ [ 1 , 4 , 1 ] } { A_1 }^{ \mp 1 } { A_2 }^{ \pm 2 } } { \sqrt{ 3 } [ 1 , 1 ] \sqrt{ [ 1 , 1 , 1 ] } } + \frac{ \sqrt{ 2 } [ 1 , - 1 ] } { \sqrt{ 5 [ 1 , 4 , 1 ] } } { A_1 }^{ \pm 1 } { A_4 }^0 \right. \nonumber \\
&\hspace{ 3mm } \left. + \frac{ [ 1 , - 1 ] [ 1 , - 8 , 1 ] { A_2 }^0 { A_3 }^{ \pm 1 } } { 5 \sqrt{ 3 } [ 1 , 1 ] [ 1 , 4 , 1 ] }
+ \frac{ 4 [ 2 , 10 , 21 , 10 , 2 ] { A_2 }^{ \pm 2 } { A_3 }^{ \mp 1 } } { 5 \sqrt{ 3 } [ 1 , 1 ] [ 1 , 4 , 1 ] \sqrt{ [ 1 , 1 , 1 ] } }
+ \frac{ [ 1 , 10 , 1 ] { A_3 }^{ \pm 1 } { A_4 }^0 } { \sqrt{ 5 } [ 1 , 4 , 1 ] } \right), \nonumber \\
{ B_3 }^{ \pm 3 } & = - \frac{ \sqrt{ 3 [ 1 , 0 , 1 ] } k^2 } { \sqrt{ [ 1, 1, 1 ] } } \left(
\frac{ { A_1 }^{ \pm 1 } { A_2 }^{ \pm 2 } }{ 2 }
+ \frac{ 2 [ 1 , - 1 ] [ 1 , 3 , 1 ] } { 5 [ 1 , 0 , 1 ] \sqrt{ [ 1 , 4 , 1 ] } }
{ A_2 }^{ \pm 2 } { A_3 }^{ \pm 1 } \right), \nonumber \\
{ B_4 }^{ 0 } & = j k { A_4 }^0 - \frac{ k^2 } { \sqrt{ 5 } } \left[
\frac{ \sqrt{ 2 } [ 1 , - 1 ] } { \sqrt{ [ 1 , 4 , 1 ] } } \left( { A_1 }^1 { A_3 }^{ -1 } + { A_1 }^{ -1 } { A_3 }^{ 1 } \right) + \frac{ [ 1, 10, 1 ] } { [ 1, 4, 1 ] } { A_3 }^1 { A_3 }^{ -1 } + \left( { A_2 }^0 \right)^2 \right. \nonumber \\
& \left. \hspace{ 3 mm } + \frac{ [ 1 , - 1 ]^2 } { 2 [ 1 , 1 , 1 ] } { A_2 }^{ 2 } { A_2 }^{ -2 } + \frac{ 5 } { 7 } \left( { A_4 }^0 \right)^2 \right], \nonumber \\
{ B_4 }^{ \pm 2 } & = - \frac{ 3 k^2 } { \sqrt{ [ 1 , 4 , 10 , 4 , 1 ] } } \left[
\frac{ [ 1 , 4 , 10 , 4 , 1 ]  { A_1 }^{ \pm 1 } { A_3 }^{ \pm 1 } } { \sqrt{ 10 [ 1 , 1 , 1 ] [ 1 , 4 , 1 ] } }
+ \frac{ [ 1 , 4 , 10 , 4 , 1 ] } { 2 \sqrt{ 5 } [ 1 , 1 , 1 ] } { A_2 }^0 { A_2 }^{ \pm 2 }
+ \frac{ [ 1 , - 1 ] [ 1 , 3, 1 ] } { 2 \sqrt{ 3 } [ 1 , 1 , 1 ] } \right. \nonumber \\
& \hspace{ 1 mm } \left. \times { A_2 }^{ \pm 2 } { A_4 }^0 - \frac{ 2 [ 0 , 0 , 1 , - 1 ] \left( { A_3 }^{ \pm 1 } \right)^2 } { \sqrt{ 5 } [ 1 , 4 , 1 ] \sqrt{ [ 1 , 1 , 1 ] } } \right], ~
{ B_4 }^{ \pm 4 } = - \frac{ 3 k^2 \sqrt{ [ 1 , 1 , 1 , 1 , 1 ] } } { 2 \sqrt{ 5 } [ 1 , 1 , 1 ] } \left( { A_2 }^{ \pm 2 } \right)^2, \label{eq:Bpq-2} \\
{ B_5 }^{ \pm 1 } & = - \frac{ \sqrt{ 3 [ 1 , 8 , 1 ] } k^2 } { \sqrt{ 5 } [ 1 , 4 , 1 ] } \left(
\sqrt{ [ 1 , 4 , 1 ] } { A_1 }^{ \pm 1 } { A_4 }^0 + \frac{ \sqrt{ 6 } [ 1 , 1 ] } { \sqrt{ 5 } } { A_2 }^0 { A_3 }^{ \pm 1 } + \frac{ \sqrt{ 3 } [ 1 , 1 ] [ 1 , - 1 ] } { \sqrt{ 10 } \sqrt{ [ 1 , 1 , 1 ] } } { A_2 }^{ \pm 2 } { A_3 }^{ \mp 1 } \right. \nonumber \\
& \hspace{ 3mm } \left.  + \frac{ \sqrt{ 2 } [ 1 , - 1 ] [ 1 , - 10 , 1 ] } { 7 [ 1 , 8 , 1 ] } { A_3 }^{ \pm 1 } { A_4 }^0 \right), ~
{ B_5 }^{ \pm 3 } = - \frac{ 3 \sqrt{ 2 } k^2 \sqrt{ [ 1, 4, 10, 20, 10, 4, 1 ] } }
{ 5 \sqrt{ [ 1, 0, 1 ] [ 1, 1, 1 ] [ 1, 4, 1 ] } } { A_2 }^{ \pm 2 } { A_3 }^{ \pm 1 }, \nonumber \\
{ B_6 }^{ 0 } & = - \frac{ 9 k^2 }{ \sqrt{ 35 } } \left( \frac{ { A_2 }^0 { A_4 }^0 } { \sqrt{ 3 } }
+ \frac{ [ 1, 1 ] [ 1, -1 ] } { \sqrt{ 5 } [ 1, 4, 1 ] } { A_3 }^1 { A_3 }^{ -1 } \right), ~
{ B_7 }^{ \pm 1 } = - \frac{ 9 \sqrt{ 10 } k^2 \sqrt{ [ 1, 16, 36, 16, 1 ] } }
{ 35 \sqrt{ [ 1, 4, 1 ] [ 1, 8, 1 ] } } { A_3 }^{ \pm 1 } { A_4 }^0, \nonumber \\
{ B_6 }^{ \pm 2 } & = - \frac{ 6 k^2 \sqrt{ [ 1, 9, 45, 65, 45, 9, 1 ] } } { \sqrt{ 35 [ 1, 4, 10, 4, 1 ] } } \left[
\frac{ { A_2 }^{ \pm 2 } { A_4 }^0 } { \sqrt{ 3 [ 1, 1, 1 ] } }
+ \frac{ [ 1, 1 ] \left( { A_3 }^{ \pm 1 } \right)^2 } { \sqrt{ 5 } [ 1, 4, 1 ] } \right], ~
{ B_8 }^{ 0 } = - \frac{ 3 k^2 } { 7 } \left( { A_4 }^0 \right)^2. \nonumber
\end{align}

\section{Decomposition of obliquely propagating Gaussian beam into Zernike annular polynomials} \label{app3}
Consider a beam propagated along an axis tilted by $ \theta_\mathrm{ bw } $ with respect to $ z $ axis in Fig.~\ref{fig1}.
The beam axis is referred to as $ z^\prime $ here and the azimuthal angle $ \phi_\mathrm{ bw } $ is defined as an angle between $ x $ axis and the projection of $ z^\prime $ axis onto $ xy $ plane.
The relation between the primed and non-primed coordinates is as follows:
\begin{align}
\begin{split}
x^\prime & = x \cos \theta_\mathrm{ bw } \cos \phi_\mathrm{ bw } + y \cos \theta_\mathrm{ bw } \sin \phi_\mathrm{ bw } - z \sin \theta_\mathrm{ bw }, \\
y^\prime & = - x \sin \phi_\mathrm{ bw } + y \cos \phi_\mathrm{ bw }, \\
z^\prime & = x \sin \theta_\mathrm{ bw } \cos \phi_\mathrm{ bw } + y \sin \theta_\mathrm{ bw } \sin \phi_\mathrm{ bw } + z \cos \theta_\mathrm{ bw }.
\end{split}
\label{eq:rotation}
\end{align}
Let us focus on the cylindrical coordinates at $ z = 0 $ in particular.
The followings hold under the assumption of $ \left| \sin \theta_\mathrm{ bw } \right| \ll 1 $, $ { r^\prime }^2 = { x^\prime }^2 + { y^\prime }^2 = { R_\mathrm{ ex } }^2 \rho^2 + O \left( \sin^2 \theta_\mathrm{ bw } \right) $, 
$ z^\prime = { R_\mathrm{ ex } } \rho \sin \theta_\mathrm{ bw } \cos \left( \psi - \phi_\mathrm{ bw } \right) $, 
$ \psi^\prime = \psi - \phi_\mathrm{ bw } + O \left( \sin^2 \theta_\mathrm{ bw } \right) $.
A Laguerre-Gaussian beam mode can be written with respect to the primed coordinate system,
\begin{align}
& { E_{ p^\prime } }^{ q^\prime } = \sqrt{ \frac{ 2 p^\prime ! } { \pi \tilde{ w }^2( z^\prime; z^\prime_\mathrm{ bw } ) \left( p^\prime + 
\left| q^\prime \right| \right) ! } } \left( \frac{ \sqrt{ 2 } r^\prime } { \tilde{ w } \left( z^\prime; z^\prime_\mathrm{ bw } \right) }
\right)^{ \left| q^\prime \right| } { L_{ p^\prime } }^{ \left| q^\prime \right| } \left( \frac{ 2 { r^\prime }^2 } { \tilde{ w }^2
\left( z^\prime; z^\prime_\mathrm{ bw } \right) } \right)
\exp \left[ - \frac{ { r^\prime }^2 } { \tilde{ w }^2 \left( z^\prime; z^\prime_\mathrm{ bw } \right) } \right] \nonumber \\
& \times \exp \left[ - j k \left( z^\prime - z^\prime_\mathrm{ bw } \right) + j \frac{ k { r^\prime }^2 }
{ 2 \tilde{ R } \left( z^\prime; z^\prime_\mathrm{ bw } \right) } + j \left( 2 p^\prime + \left| q^\prime \right| + 1 \right)
\tilde{ \phi }_0 \left( z^\prime; z^\prime_\mathrm{ bw } \right) + j q^\prime \psi^\prime \right],
\label{eq:Epq_tilted}
\end{align}
where each parameter is given in Section~\ref{sec2.5}.
Since the beam waist is at $ ( x_\mathrm{ bw }, y_\mathrm{ bw }, z_\mathrm{ bw } ) = ( z^\prime_\mathrm{ bw } \sin \theta_\mathrm{ bw } \cos \phi_\mathrm{ bw }, z^\prime_\mathrm{ bw } \sin \theta_\mathrm{ bw }
\sin \phi_\mathrm{ bw }, z^\prime_\mathrm{ bw } \cos \theta_\mathrm{ bw } )$ and the conversion between the primed and non-primed coordinate systems is defined in Eq.~(\ref{eq:rotation}),
the parameters are expressed with respect to the non-primed coordinate system,
\begin{align}
\begin{split}
\frac{ 1 } { \tilde{ w }^2 \left( z^\prime; z^\prime_\mathrm{ bw } \right) }
& \sim \frac{ T_\mathrm{ e } } { 2 { R_\mathrm{ ex } }^2 } \left( 1 + \frac{ 4 T_\mathrm{ e } z_\mathrm{ bw } \rho
\sin \theta_\mathrm{ bw } \cos \left( \psi - \phi_\mathrm{ bw } \right) } { k^2 { w_\mathrm{ bw } }^2 R_\mathrm{ ex } } \right), \\
\frac{ { r^\prime }^2 } { \tilde{ R } \left( z^\prime; z^\prime_\mathrm{ bw } \right) }
& \sim \frac{ { R_\mathrm{ ex } }^2 \rho^2 } { R \left( 0; z_\mathrm{ bw } \right) } \left[ 1 - \frac{ { R_\mathrm{ ex } } \rho \sin \theta_\mathrm{ bw } \cos \left( \psi - \phi_\mathrm{ bw } \right) } { z_\mathrm{ bw } } \left( 1 - \frac{ 4 T_\mathrm{ e } z_\mathrm{ bw }^2 } { k^2 { w_\mathrm{ bw } }^2 { R_\mathrm{ ex } }^2 } \right) \right], \\
\tilde{ \phi }_0 ( z^\prime; z^\prime_\mathrm{ bw } ) 
& \sim \phi_0 ( 0; z_\mathrm{ bw } ) + \frac{ T_\mathrm{ e } \rho \sin \theta_\mathrm{ bw }
\cos \left( \psi - \phi_\mathrm{ bw } \right) } { k R_\mathrm{ ex } }.
\end{split}
\label{eq:rotated_beam_parameters}
\end{align}
After some manipulation with Eq.~(\ref{eq:e_det_z}), (\ref{eq:zap_norm}), (\ref{eq:Epq_tilted}), (\ref{eq:rotated_beam_parameters}), we finally obtain
\begin{align}
{ C_p }^q \left( \boldsymbol{ r }_\mathrm{ bw }; \boldsymbol{ r }_\mathrm{ ref }; w_\mathrm{ bw }; \varepsilon \right) = \sum_{ p^\prime, q^\prime } { D_{ p^\prime } }^{ q^\prime } \int_\varepsilon^1 \rho \, \mathrm{ d } \rho \int_0^{ 2 \pi } \mathrm{ d } \psi \frac{ R_\mathrm{ ex } { E_{ p^\prime } }^{ q^\prime } { Z_p^\ast }^q } { \pi [ 1, - 1 ] E_\mathrm{ sph } }, \label{eq:app_c}
\end{align}
where
\begin{align}
& \frac{ { E_{ p^\prime } }^{ q^\prime } R_\mathrm{ ex } } { E_\mathrm{ sph } } = \sqrt{ \frac{ { T_\mathrm{ e } }^{ | q^\prime | + 1 }
p^\prime ! } { \pi \left( p^\prime + \left| q^\prime \right| \right) ! } }
\exp \left[ j k z_\mathrm{ bw } + j \left( 2 p^\prime + \left| q^\prime \right| + 1 \right) \phi_0 - \frac{ T_\mathrm{ e } } { 2 } \rho^2 + j q^\prime \left( \psi - \phi_\mathrm{ bw } \right) \right] \rho^{ \left| q^\prime \right| } \nonumber \\
& \times \left\{  { L_{ p^\prime } }^{ \left| q^\prime \right| } \left( T_\mathrm{ e } \rho^2 \right) \left[ 1 + j k R_\mathrm{ ex } \rho \left( \sin \theta_\mathrm{ ref } \cos \left( \psi - \phi_\mathrm{ ref } \right)
- \sin \theta_\mathrm{ bw } \cos \left( \psi - \phi_\mathrm{ bw } \right) \right) + \frac{ j k { R_\mathrm{ ex } }^2 \rho^2 } { 2 } \right. \right. \nonumber \\
& \hspace{ 6 mm } \times \left( \frac{ 1 } { R \left( 0; z_\mathrm{ bw } \right) } - \frac{ 1 } { z_\mathrm{ ref } } \right) + \frac{ T_\mathrm{ e } \rho \sin \theta_\mathrm{ bw } \cos \left( \psi - \phi_\mathrm{ bw } \right) }
{ k R_\mathrm{ ex } } \left( \frac{ 2 \left( 1 + \left| q^\prime \right| \right) z_\mathrm{ bw } } { k { w_\mathrm{ bw } }^2 } + j \left( 2 p^\prime + \left| q^\prime \right| + 1 \right) \right) \nonumber \\
& \left. \hspace{ 6 mm } - \frac{ 2 { T_\mathrm{ e } }^2 z_\mathrm{ bw } \rho^3 \sin \theta_\mathrm{ bw } \cos \left( \psi - \phi_\mathrm{ bw } \right) } { k^2 { w_\mathrm{ bw } }^2 R_\mathrm{ ex } } \left( 1 - j \frac{ 2 z_\mathrm{ bw } } { k { w_\mathrm{ bw } }^2 } \left( 1 - \frac{ k^2 { w_\mathrm{ bw } }^2 { R_\mathrm{ ex } }^2 } { 4 T_\mathrm{ e } { z_\mathrm{ bw } }^2 } \right) \right) \right] \nonumber \\
& \left. - { L_{ p^\prime - 1 } }^{ \left| q^\prime \right| + 1 } \left( T_\mathrm{ e } \rho^2 \right) \frac{ 4 z_\mathrm{ bw } { T_\mathrm{ e } }^2 \rho^3 \sin \theta_\mathrm{ bw } \cos \left( \psi - \phi_\mathrm{ bw } \right) } { k^2 { w_\mathrm{ bw } }^2 R_\mathrm{ ex } } \right\}.
\label{eq:tilted_field}
\end{align}
Eq.~(\ref{eq:tilted_field}) is the result of the expansion of $ { E_{ p^\prime } }^{ q^\prime } $ into the Taylor series assuming $ R ( 0; z_\mathrm{ bw } ) \approx z_\mathrm{ ref } $, $ \left| \sin \theta_\mathrm{ ref } \right| \ll 1 $, and  $ \left| \sin \theta_\mathrm{ bw } \right| \ll 1 $.

\section{Polynomials ${\tilde{R}_n}{}^{|m|}(I_p)$} \label{app4}
Integral $I_p$ is defined in Eq.~(\ref{eq:Ip_def}).
Polynomials $ { \tilde{ R }_n } {}^{ | m | } ( I_p ) $ is obtained from replacing $ \rho^n, \rho^{ n - 2 }, \cdots, \rho^{|m|} $ with the integrals $ I_p, I_{ p - 2 }, \cdots,  I_{ p - n + \left| m \right| }$, respectively.
Take an example (cf. Appendix~\ref{app1}),
\begin{align}
{ \tilde{ R }_5 } {}^{ 1 } ( I_p ) &= \cfrac{ 10 [ 1, 4, 1 ] I_p - 12 [ 1, 4, 4, 1 ] I_{ p - 2 } + 3 [ 1, 4, 10, 4, 1 ] I_{ p - 4 } } { [ 1, - 1 ] ^2\sqrt{ [ 1, 4, 1 ] [ 1, 9, 9, 1 ] } }.
\label{eq:Rtilde}
\end{align}

\section*{Funding}
Grant-in-Aid for JSPS Fellows 13J01017 (HI), MEXT KAKENHI Grant Number 15K17598(MN).

\section*{Acknowledgements}
We thank the National Institute of Information and Communications Technology for supporting the PO simulation and to Professor Naomasa Nakai at University of Tsukuba for supporting the ray tracing simulation.
We are grateful to Alvaro Gonzalez at the National Astronomical Observatory of Japan and Yutaro Sekimoto at Institute of Space and Astronautical Science, Japan Aerospace Exploration Agency for the fruitful discussions.
We appreciate the reviewer's comment which significantly improves the present paper.

\section*{Disclosures}
The authors declare no conflicts of interest.

\bibliography{ref}

\end{document}